\documentclass[twocolumn]{aastex63}

\providecommand{\gaia}{{\it Gaia}}

\providecommand{\parallax}{\ensuremath{\varpi}}
\providecommand{\sigparallax}{\ensuremath{\sigma_{\varpi}}}

\providecommand{\glon}{\ensuremath{l}}
\providecommand{\glat}{\ensuremath{b}}

\providecommand{\dist}{\ensuremath{r}}

\shortauthors{Zhang et al.}

\begin{document}

\title{A New Transition Wolf-Rayet WN/C Star in the Milky Way}

\correspondingauthor{Wei Zhang}
\email{xtwfn@bao.ac.cn}

\author{Wei Zhang}
\affiliation{CAS Key Laboratory of Optical Astronomy, National Astronomical Observatories,  Chinese Academy of Sciences, Beijing 100101, People's Republic of China}

\author{Helge Todt}
\affiliation{Institut f\"ur Physik und Astronomie, Universit\"at Potsdam, Karl-Liebknecht-Str. 24/25, 14476 Potsdam, Germany}

\author{Hong Wu}
\affiliation{CAS Key Laboratory of Optical Astronomy, National Astronomical Observatories,  Chinese Academy of Sciences, Beijing 100101, People's Republic of China}

\author{Jianrong Shi}
\affiliation{CAS Key Laboratory of Optical Astronomy, National Astronomical Observatories,  Chinese Academy of Sciences, Beijing 100101, People's Republic of China}

\author{Chih-Hao Hsia}
\affiliation{State Key Laboratory of Lunar and Planetary Sciences, Macau University of Science and Technology, Taipa, Macau, People's Republic of China}

\author{Yuzhong Wu}
\affiliation{CAS Key Laboratory of Optical Astronomy, National Astronomical Observatories,  Chinese Academy of Sciences, Beijing 100101, People's Republic of China}

\author{Chaojian Wu}
\affiliation{CAS Key Laboratory of Optical Astronomy, National Astronomical Observatories,  Chinese Academy of Sciences, Beijing 100101, People's Republic of China}

\author{Yongheng Zhao}
\affiliation{CAS Key Laboratory of Optical Astronomy, National Astronomical Observatories,  Chinese Academy of Sciences, Beijing 100101, People's Republic of China}

\author{Tianmeng Zhang}
\affiliation{CAS Key Laboratory of Optical Astronomy, National Astronomical Observatories,  Chinese Academy of Sciences, Beijing 100101, People's Republic of China}

\author{Yonghui Hou}
\affiliation{University of Chinese Academy of Sciences, Beijing 100049, People's Republic of China}
\affiliation{Nanjing Institute of Astronomical Optics, \& Technology, National Astronomical Observatories, Chinese Academy of Sciences, Nanjing 210042, People's Republic of China}

        
\label{firstpage}

\begin{abstract}

We report the discovery of a new transition type Wolf-Rayet (WR) WN/C star in
the Galaxy.  According to its coordinates (R.A., Dec)$_{\rm J2000}$ = $\rm
18^h51^m39\fs7$, $-05\arcdeg34\arcmin51\farcs1$, and the distance
(7.11$^{+1.56}_{-1.22}\,$kpc away from Earth) inferred from the second \gaia\,
data release, it's found that WR 121-16 is located in the Far 3\,kpc Arm, and it
is 3.75\,kpc away from the Galactic Center.  The optical spectra obtained by the
Large Sky Area Multi-Object Fiber Spectroscopic Telescope (LAMOST) and the
2.16\,m telescope, both located at the Xinglong Observatory in China, indicate that this
is a WR star of the transitional WN7o/WC subtype.  A current stellar mass
of about 7.1$^{+1.7}_{-1.1}\,$M$_\sun$, a mass-loss rate of
$\dot{M}=10^{-4.97^{+0.16}_{-0.20}}\,M_\sun\,\rm yr^{-1}$, a bolometric
luminosity of log $L/L_\sun$ = 4.88$^{+0.17}_{-0.15}$, and a stellar
temperature of $T_*=47^{+9}_{-5}\,$kK are derived, by fitting the observed spectrum with a
specific Potsdam Wolf-Rayet (PoWR) model.  The magnitude in $V$-band varies
between 13.95 and 14.14 mag, while no period is found. Based on the optical
spectra, the time domain data, and the indices of the astrometric solution of the \gaia\ data, WR
121-16 is likely a transitional WN/C single star rather than a WN+WC binary.

\end{abstract}
       

\section{Introduction}

Wolf-Rayet (WR) stars are evolved descendants of the massive O stars with
initial mass $M_{\rm i}\ga 25\,M_\sun$. Unlike most stars whose spectra show narrow
absorption lines, the WR stars have spectacular spectra with strong and broad
emission lines, which are formed in the optically thick and extremely fast
stellar winds \citep[e.g.,][]{Crowther-07}. With a high mass-loss rate ($\sim
10^{-5}-10^{-4}\,M_\sun\,\rm yr^{-1}$) and ending their lives as core-collapse
supernovae, they play an important role in shaping the structure and chemical
evolution of their host galaxies. WR stars can also be progenitors of
long-duration gamma-ray bursts \citep{Callingham-Tuthill-Pope-19}.  The number
of WR stars in the Milky Way is expected to be 1200 $\pm$ 200
\citep{Rosslowe-Crowther-15a,Rosslowe-Crowther-15b}. Even with the help of
infrared observations which are suitable to detect obscured targets
\citep{Hillier-Jones-Hyland-83,Hillier-85,Crowther-Smith-96,Roman-Lopes-20},
only 666 WR stars have been found to
date\footnote{http://pacrowther.staff.shef.ac.uk/WRcat/index.php, v1.24},
indicating that about half of them are still hidden in the Galaxy.

\begin{figure*}
\centering
\includegraphics[width=\textwidth]{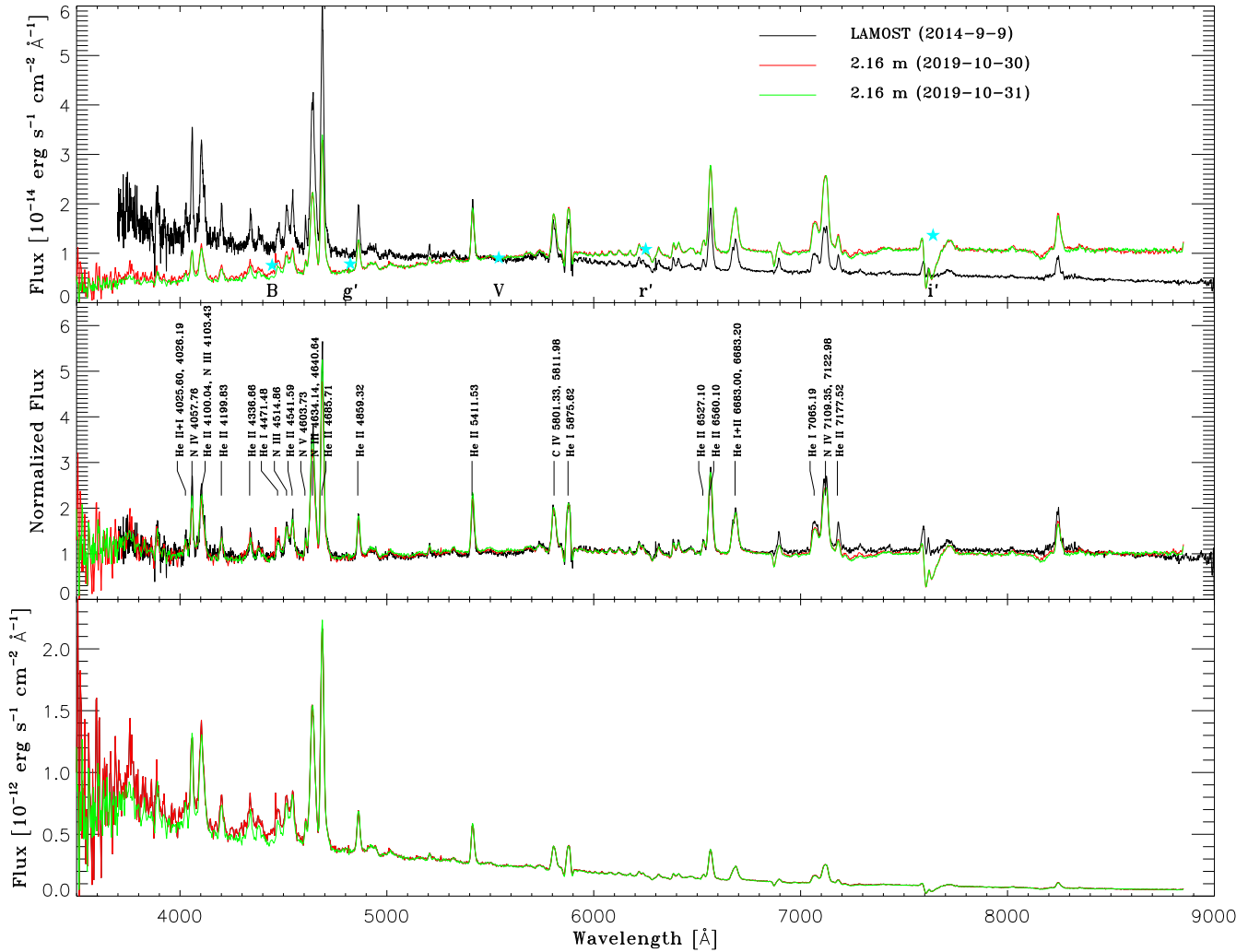}
\caption{The observed spectra obtained at the LAMOST (black lines) and the 2.16\,m telescope (red and green lines). In the top panel, the LAMOST spectrum has been shifted to a visual $V$ magnitude of 14.024. The SED is shown as cyan filled stars. These three continuum normalized spectra with some line identifications are shown in the middle panel. The dereddened 2.16\,m spectra are shown in the bottom panel. [The data of the three spectra are available in the electronic edition of the Journal.]}
\label{fig:spectra}
\end{figure*}

According to the strong emission lines, WR stars can be classified as WN
(helium and nitrogen), WC and WO (helium, carbon and oxygen) types.  WN stars
are believed to show the hydrogen burning products via the CNO cycle, while WC
stars reveal the helium burning products via the triple-$\alpha$ cycle.  Based
on the strength of the emission lines and line ratios, WN stars can be further
classified into the spectral subtypes WN2 to WN11, and WC stars into the
spectral subtypes WC4 to WC9 \citep{Smith-68, Smith-Shara-Moffat-90,
Smith-Crowther-Prinja-94, Smith-Shara-Moffat-96, Crowther-DeMarco-Barlow-98}.
Also, there are transition types from WN to WC, which are called WN/C or WC/N
stars, whose spectra show strong emission lines of carbon and nitrogen
simultaneously \citep{Massey-Grove-89, Conti-Massey-89, Crowther-07}. As the
duration of this transition phase is very short ($\sim$ 1 Myr), they are hard to
find \citep{Crowther-Smith-Willis-95}. Only about a dozen of transition type
stars are currently known in the Milky Way, the LMC, the SMC, IC10, NGC 1313,
M31, M33 and M81 \citep{Morgan-Good-87,Conti-Massey-89,Schild-Smith-Willis-90,
Breysacher-Azzopardi-Testor-99,Crowther-03, Hadfield-Crowther-07,
Massey-14,Shara-16,Gomez-Gonzalez-Mayya-Rosa-Gonzalez-16}. We note that no
WN/C star in M31 has been known to us until the first one was detected by
\cite{Shara-16}.

Having a large field of view (5\arcdeg\ in diameter) and taking 4000 spectra in
a single exposure, the Large Sky Area Multi-Object Fiber Spectroscopic
Telescope (LAMOST) at the Xinglong Observatory in China is powerful to search
for such rare objects. The strong emission lines of our target can be easily
identified in the low-resolution spectrum (R$\sim$1800), which covers the full
optical range from 3700 to 9000 \AA\ \citep{Wang-96, Su-Cui-04,
Cui-12,Zhao-12,Luo-15}.  We here report a new WN/C star serendipitously
discovered during the LAMOST testing observations carried out in the full moon
nights.

The paper is structured as follows. In Section 2,  we describe
the observation and reduction of the optical spectra, and the spectral
classification of the newly discovered WR star.  The stellar parameters are
given in Section 3. We discuss the probability of the new WR star being a binary in Section 4, 
and summarize our results in Section 5.

\section{data Observation and analysis}

\subsection{Optical Spectra} 

The WR star discussed in this paper was found as a `by-product' of the LAMOST
testing observations during the full moon nights, when the telescope was
pointing to the open cluster M11. The equatorial and galactic coordinates of
this object are (R.A., Dec) = ($\rm 18^h51^m39\fs7$,
$-05\arcdeg34\arcmin51\farcs1$) and (\glon, \glat) = ($\rm 01^h51^m56\fs9$,
$-02\arcdeg35\arcmin28\farcs0$), respectively, which is about 42\farcm24 apart
from the center of M11.  Following the revised nomenclature scheme introduced
in Appendix A of \cite{Rosslowe-Crowther-15a} for Galactic WR stars, we name it
WR 121-16.  The visual magnitude in $V$-band for this star is 14.0, and the
spectral energy distribution (SED) from the optical to the infrared is
summarized in Table \ref{tbl:sed}. The spectrum obtained on September 9, 2014,
with an exposure time of 1800\,s,  is shown in the top panel of Figure
\ref{fig:spectra} (black line).  The data was reduced using the standard
pipeline, including bias subtraction, flat correction, spectra extraction,
wavelength calibration, and flux calibration \citep{Luo-15}.  This 
low-resolution (R$\sim$1800) spectrum shows strong nitrogen lines, such as
N{\,\sc iii} $\lambda$4634-$\lambda$4641 blend, N{\,\sc iv} $\lambda$4057,
$\lambda$7109, $\lambda$7123. The N{\,\sc v} $\lambda$4604,
$\lambda$4933-$\lambda$4944 blend lines are relatively weak but can be clearly
detected. The N{\,\sc v} $\lambda$4620 line is weak and blended to
$\lambda$4634-$\lambda$4641 blend lines. The He{\,\sc i} and He{\,\sc ii} lines
are also strong.  The C{\,\sc iv} $ \lambda$5808 line is strong, while C{\,\sc iii}
$\lambda$5696 line can not be detected.

\begin{table}
\tiny
\begin{center}
  \caption{The SED of the new WN/C star.}
    \begin{tabular}{llclclr}\hline\hline
Filters  & Catalog   & freqency (GHZ)        & mag       &   mag err & magtype & $\Delta$mag$^a$\\
\hline  
\hline  
B       & APASS9   &   674.90e+3       & 14.819     &   0.017   &   Vega    & -0.163  \\
V       & APASS9   &   541.43e+3       & 14.024     &   ---     &   Vega    & -0.044  \\
g$^\prime$      & APASS9   &   621.98e+3       & 14.436     &   ---     &   asinh   &  0.000  \\
r$^\prime$      & APASS9   &   479.90e+3       & 13.528     &   ---     &   asinh   &  0.000  \\
i$^\prime$      & APASS9   &   392.66e+3       & 12.840     &   0.033   &   asinh   &  0.000  \\
J       & 2MASS    &   241.96e+3       & 11.369     &   0.032   &   Vega    &  0.910  \\
H       & 2MASS    &   181.75e+3       & 10.997     &   0.022   &   Vega    &  1.380  \\
Ks      & 2MASS    &   138.55e+3       & 10.609     &   0.019   &   Vega    &  1.850  \\
W1      & WISE     &   89.490e+3       & 9.845      &   0.137   &   Vega    &  2.699  \\
W2      & WISE     &   65.172e+3       & 9.747      &   0.115   &   Vega    &  3.339  \\
W3      & WISE     &   25.934e+3       & 8.562      &   0.139   &   Vega    &  5.174  \\
W4      & WISE     &   13.571e+3       & 7.505      &   0.161   &   Vega    &  6.620  \\
\hline  
    \end{tabular}
  \label{tbl:sed}
\end{center}
\tablecomments{~$^a$ mag(AB) = mag + $\Delta$mag}
\end{table}

We also carried out the followup observations using the BAO Faint Object
Spectrograph and Camera (BFOSC) on the 2.16\,m telescope at the Xinglong
Observatory in China on October 30 and 31, 2019, with an exposure time of 1200\,s
on each night \citep{Zhao-18}.  The grating 300 l mm$^{-1}$, blazed at 6000
\AA, and a slit width of 2.\arcsec0 was used and provided spectral coverage
roughly from 3250 to 8850 \AA\ with a resolving power R$\sim$1600.  The
data was reduced using standard IRAF procedures, and the spectra are absolutely
flux-calibrated using the standard star HR 7596, which was observed in the same
night with an exposure time of 5\,s.  These two spectra (red and blue lines)
have been compared with the LAMOST spectrum (black line) in Figure
\ref{fig:spectra}. In the top panel, the LAMOST spectrum has been shifted to
$V$ = 14.024 mag to make a comparison with the 2.16\,m spectra. We can see that the
2.16\,m spectra are identical with each other, while they have quite different
slopes of the continuum compared to that one of the LAMOST spectrum.  As the 2.16\,m
spectra have been absolutely flux-calibrated, we then compare them with the
observed SED which is summarized Table \ref{tbl:sed}. Note that the SDSS
asinh magnitudes in g$^\prime$, r$^\prime$ and i$^\prime$-band are close to the AB system, while the Vega
magnitudes should be converted to the AB system by adding the offsets $\rm \Delta
mag$ listed in the last column of Table \ref{tbl:sed}, and the data of $\rm \Delta
mag$ are from \cite{Frei-Gunn-94}, \cite{Blanton-07} and \cite{Jarrett-11}. It
is found that the 2.16\,m spectra are consistent with the SED shown as cyan
stars.  Hereafter, it tell us that there was some problem with the relative
flux calibration for the LAMOST observation.  This result is not strange for the large
field of view and multi-fiber observations in the direction of high extinction
regions. We have normalized all the three spectra and present them in the
middle panel of Figure \ref{fig:spectra}.  We can see the line strengths of the
LAMOST spectrum are now almost identical with those of the 2.16\,m spectra.

We have measured the most prominent emission features, including the observed
central wavelength, full width at half maximum (FWHM), flux and equivalent
width (EW), and listed the results in Table \ref{tbl:lines}. We should note
that the flux of the LAMOST spectrum is not listed in this table. Moreover,
as the spectra are affected by strong reddening (see Section 3.2), the measured
fluxes from the original fluxed spectra may not represent the star properties.
We then also measured the fluxes from the dereddened spectra (see the bottom panel of Figure \ref{fig:spectra}) and listed them
below the original ones. From this table, we can see that these
lines do not change with time, as shown in the middle panel of Figure
\ref{fig:spectra}.

\begin{longrotatetable}
\begin{deluxetable*}{rcccccccccccccccc}
\tablecaption{Observational properties of the prominent emission lines. Flux is in units of $\rm 10^{-14}~erg~s^{-1}~cm^{-2}$. For each emission line, the first row 
lists measurement results from the original fluxed spectrum, while the second row lists that from the dereddened one. \label{tbl:lines}}
\tablewidth{700pt}
\tabletypesize{\tiny}
\tablehead{   
          \vline & \multicolumn{4}{c}{LAMOST (2014-9-9)} \vline& \multicolumn{4}{c}{2.16\, m (2019-10-30)} \vline &  \multicolumn{4}{c}{2.16\,m (2019-10-31)} \\
\hline
\colhead{Line Name} &
\colhead{Wave} & \colhead{FWHM} & \colhead{Flux} & \colhead{EW} & 
\colhead{Wave} & \colhead{FWHM} & \colhead{Flux} & \colhead{EW} & 
\colhead{Wave} & \colhead{FWHM} & \colhead{Flux} & \colhead{EW}\\
          &
\colhead{(\AA)} & \colhead{(\AA)} & ($^a$) & \colhead{(\AA)} & 
\colhead{(\AA)} & \colhead{(\AA)} & ($^a$) & \colhead{(\AA)} & 
\colhead{(\AA)} & \colhead{(\AA)} & ($^a$) & \colhead{(\AA)}
}             
\startdata 
             He{\,\sc ii+i} 4025.60, 4026.19 & 4028.85 $\pm$    4.35  & 11.3 $\pm$  10.8 &        --          &   4.6 $\pm$   4.2 &  4026.56 $\pm$   64.04 &  30.7 $\pm$ 193.0  &  4.1 $\pm$  27.3 &   8.7 $\pm$  58.0 &  4027.22 $\pm$   27.02 &  12.2 $\pm$  68.2  &  1.9 $\pm$  10.6 &   4.3 $\pm$  23.8 \\ 
               & & & & & 4026.37 $\pm$   0.52 &  32.1 $\pm$  1.6 &   547.1 $\pm$   29.5 &   9.2 $\pm$  0.5  &  4027.19 $\pm$   0.21 &  12.2 $\pm$  0.5 &   246.2 $\pm$   10.6 &   4.3 $\pm$  0.2 \\ 
                  N{\,\sc iv} 4057.76 & 4058.32 $\pm$    1.02  & 10.4 $\pm$   2.5 &        --          &  17.5 $\pm$   4.0 &  4057.93 $\pm$   10.44 &  15.9 $\pm$  24.1  &  9.3 $\pm$  15.1 &  19.3 $\pm$  31.3 &  4058.03 $\pm$    7.40 &  15.1 $\pm$  18.6  &  9.7 $\pm$  11.9 &  21.4 $\pm$  26.3 \\ 
               & & & & & 4057.91 $\pm$   0.08 &  15.9 $\pm$  0.2 &  1149.7 $\pm$   15.5 &  19.3 $\pm$  0.3  &  4057.99 $\pm$   0.06 &  15.1 $\pm$  0.2 &  1196.9 $\pm$   11.9 &  21.3 $\pm$  0.2 \\ 
  He{\,\sc ii} 4100.04, N{\,\sc iii} 4103.43 & 4105.53 $\pm$    1.93  & 23.2 $\pm$   5.3 &        --          &  31.3 $\pm$   7.8 &  4103.47 $\pm$    9.80 &  26.5 $\pm$  27.9  & 17.4 $\pm$  20.9 &  34.8 $\pm$  41.9 &  4104.58 $\pm$   10.17 &  26.6 $\pm$  29.2  & 16.7 $\pm$  21.1 &  36.0 $\pm$  45.4 \\ 
               & & & & & 4103.33 $\pm$   0.08 &  26.6 $\pm$  0.2 &  2065.4 $\pm$   21.1 &  35.0 $\pm$  0.4  &  4104.45 $\pm$   0.09 &  26.5 $\pm$  0.2 &  1976.7 $\pm$   21.0 &  35.8 $\pm$  0.4 \\ 
        He{\,\sc ii} 4199.83 & 4201.34 $\pm$    2.94  &  9.5 $\pm$   7.2 &        --          &   5.8 $\pm$   4.1 &  4200.59 $\pm$   19.15 &  17.3 $\pm$  48.8  &  4.6 $\pm$  13.2 &   9.4 $\pm$  26.8 &  4201.55 $\pm$   21.04 &  17.5 $\pm$  53.7  &  4.3 $\pm$  13.3 &   9.5 $\pm$  29.3 \\ 
      & & & & & 4200.53 $\pm$   0.18 &  17.2 $\pm$  0.5 &   498.9 $\pm$   13.1 &   9.4 $\pm$  0.2  &  4201.50 $\pm$   0.19 &  17.5 $\pm$  0.5 &   464.4 $\pm$   13.3 &   9.5 $\pm$  0.3 \\ 
                 He{\,\sc ii} 4336.66 & 4341.65 $\pm$    3.70  & 11.7 $\pm$   9.2 &        --          &   6.9 $\pm$   5.2 &  4340.35 $\pm$   25.74 &  23.5 $\pm$  65.8  &  6.3 $\pm$  17.9 &  11.2 $\pm$  32.0 &  4339.15 $\pm$   24.39 &  20.0 $\pm$  63.5  &  4.5 $\pm$  14.9 &   9.2 $\pm$  30.3 \\ 
              & & & & & 4340.25 $\pm$   0.27 &  23.5 $\pm$  0.7 &   586.2 $\pm$   17.8 &  11.1 $\pm$  0.3  &  4339.06 $\pm$   0.26 &  20.2 $\pm$  0.7 &   430.4 $\pm$   15.0 &   9.3 $\pm$  0.3 \\ 
                  He{\,\sc i} 4471.48 & 4476.09 $\pm$    6.81  & 15.8 $\pm$  17.3 &        --          &   6.1 $\pm$   6.6 &  4472.00 $\pm$   24.10 &  27.6 $\pm$  69.5  &  7.8 $\pm$  21.8 &  12.9 $\pm$  36.4 &  4475.91 $\pm$   26.52 &  20.7 $\pm$  70.0  &  4.4 $\pm$  15.4 &   8.4 $\pm$  28.9 \\ 
                & & & & & 4471.88 $\pm$   0.29 &  27.6 $\pm$  0.8 &   643.5 $\pm$   21.9 &  13.0 $\pm$  0.4  &  4475.85 $\pm$   0.32 &  20.8 $\pm$  0.8 &   371.1 $\pm$   15.5 &   8.5 $\pm$  0.4 \\ 
                 N{\,\sc iii} 4514.86 & 4517.31 $\pm$    4.15  & 17.6 $\pm$  10.7 &        --          &  12.2 $\pm$   7.0 &  4515.72 $\pm$   18.57 &  22.5 $\pm$  49.5  &  9.7 $\pm$  19.3 &  15.7 $\pm$  31.3 &  4514.95 $\pm$   18.61 &  21.5 $\pm$  48.8  &  8.6 $\pm$  17.7 &  15.3 $\pm$  31.4 \\ 
               & & & & & 4515.61 $\pm$   0.23 &  22.4 $\pm$  0.6 &   766.1 $\pm$   19.3 &  15.7 $\pm$  0.4  &  4514.86 $\pm$   0.24 &  21.6 $\pm$  0.6 &   690.9 $\pm$   17.9 &  15.6 $\pm$  0.4 \\ 
                 He{\,\sc ii} 4541.59 & 4544.26 $\pm$    3.18  & 15.0 $\pm$   8.2 &        --          &  12.6 $\pm$   6.6 &  4544.30 $\pm$   15.83 &  19.7 $\pm$  39.5  &  9.5 $\pm$  19.8 &  15.2 $\pm$  31.5 &  4543.21 $\pm$   14.87 &  19.3 $\pm$  37.7  &  9.4 $\pm$  18.5 &  16.1 $\pm$  31.7 \\ 
               & & & & & 4544.22 $\pm$   0.21 &  19.8 $\pm$  0.5 &   734.3 $\pm$   19.9 &  15.2 $\pm$  0.4  &  4543.16 $\pm$   0.19 &  19.3 $\pm$  0.5 &   720.5 $\pm$   18.6 &  16.1 $\pm$  0.4 \\ 
                   N{\,\sc v} 4603.73 & 4607.31 $\pm$    3.19  &  5.8 $\pm$   7.7 &        --          &   2.8 $\pm$   3.4 &  4607.39 $\pm$   15.47 &   6.9 $\pm$  37.4  &  1.4 $\pm$   7.1 &   2.1 $\pm$  10.5 &  4607.43 $\pm$   16.80 &   8.2 $\pm$  40.8  &  1.7 $\pm$   7.9 &   2.6 $\pm$  12.2 \\ 
                 & & & & & 4607.38 $\pm$   0.22 &   6.9 $\pm$  0.5 &   102.2 $\pm$    7.1 &   2.1 $\pm$  0.1  &  4607.43 $\pm$   0.23 &   8.2 $\pm$  0.6 &   122.4 $\pm$    7.9 &   2.6 $\pm$  0.2 \\ 
        N{\,\sc iii} 4634.14, 4640.64 & 4641.70 $\pm$    1.20  & 23.6 $\pm$   3.0 &        --          &  61.6 $\pm$   8.0 &  4640.92 $\pm$    3.76 &  24.6 $\pm$   9.6  & 40.0 $\pm$  15.9 &  58.0 $\pm$  23.1 &  4641.20 $\pm$    3.65 &  25.0 $\pm$   9.3  & 42.2 $\pm$  16.0 &  63.8 $\pm$  24.2 \\ 
      & & & & & 4640.80 $\pm$   0.05 &  24.6 $\pm$  0.1 &  2766.0 $\pm$   15.9 &  58.0 $\pm$  0.3  &  4641.08 $\pm$   0.05 &  25.0 $\pm$  0.1 &  2922.6 $\pm$   16.0 &  63.8 $\pm$  0.3 \\ 
                 He{\,\sc ii} 4685.71 & 4688.22 $\pm$    0.61  & 15.9 $\pm$   1.6 &        --          &  67.0 $\pm$   6.7 &  4687.77 $\pm$    1.88 &  17.9 $\pm$   5.0  & 49.5 $\pm$  14.3 &  70.2 $\pm$  20.3 &  4688.26 $\pm$    1.78 &  17.7 $\pm$   4.7  & 51.1 $\pm$  14.0 &  74.9 $\pm$  20.6 \\ 
               & & & & & 4687.70 $\pm$   0.03 &  17.9 $\pm$  0.1 &  3259.2 $\pm$   14.3 &  70.0 $\pm$  0.3  &  4688.19 $\pm$   0.03 &  17.7 $\pm$  0.1 &  3357.5 $\pm$   14.0 &  74.7 $\pm$  0.3 \\ 
                 He{\,\sc ii} 4859.32 & 4862.82 $\pm$    2.85  & 13.2 $\pm$   7.1 &        --          &  13.4 $\pm$   7.0 &  4862.77 $\pm$    8.10 &  16.4 $\pm$  20.6  & 10.2 $\pm$  12.7 &  15.0 $\pm$  18.8 &  4863.07 $\pm$    7.62 &  16.1 $\pm$  19.4  & 10.5 $\pm$  12.6 &  15.5 $\pm$  18.6 \\ 
               & & & & & 4862.72 $\pm$   0.15 &  16.5 $\pm$  0.4 &   553.0 $\pm$   12.7 &  15.0 $\pm$  0.3  &  4863.02 $\pm$   0.14 &  16.1 $\pm$  0.4 &   568.5 $\pm$   12.6 &  15.4 $\pm$  0.3 \\ 
                 He{\,\sc ii} 5411.53 & 5414.87 $\pm$    2.87  & 15.5 $\pm$   7.3 &        --          &  19.6 $\pm$   9.1 &  5415.18 $\pm$    4.75 &  17.1 $\pm$  12.1  & 18.4 $\pm$  13.1 &  21.1 $\pm$  15.0 &  5415.28 $\pm$    4.58 &  17.3 $\pm$  11.7  & 19.5 $\pm$  13.3 &  22.6 $\pm$  15.4 \\ 
               & & & & & 5415.13 $\pm$   0.15 &  17.1 $\pm$  0.4 &   565.1 $\pm$   13.1 &  21.0 $\pm$  0.5  &  5415.23 $\pm$   0.15 &  17.3 $\pm$  0.4 &   598.8 $\pm$   13.3 &  22.5 $\pm$  0.5 \\ 
         C{\,\sc iv} 5801.33, 5811.98 & 5807.15 $\pm$    5.34  & 22.1 $\pm$  14.0 &        --          &  19.2 $\pm$  12.9 &  5808.06 $\pm$    7.06 &  24.0 $\pm$  18.8  & 21.3 $\pm$  18.2 &  22.1 $\pm$  18.8 &  5808.51 $\pm$    6.95 &  24.6 $\pm$  18.5  & 22.5 $\pm$  18.4 &  23.9 $\pm$  19.5 \\ 
      & & & & & 5807.97 $\pm$   0.31 &  23.9 $\pm$  0.8 &   475.2 $\pm$   18.1 &  21.9 $\pm$  0.8  &  5808.42 $\pm$   0.31 &  24.5 $\pm$  0.8 &   500.7 $\pm$   18.3 &  23.7 $\pm$  0.9 \\ 
                  He{\,\sc i} 5875.62 & 5877.34 $\pm$    4.84  & 21.5 $\pm$   6.9 &        --          &  40.2 $\pm$  28.5 &  5877.26 $\pm$    6.20 &  22.2 $\pm$   8.5  & 42.3 $\pm$  47.8 &  38.9 $\pm$  44.0 &  5877.58 $\pm$    6.13 &  21.8 $\pm$   8.3  & 40.6 $\pm$  53.0 &  37.9 $\pm$  49.4 \\ 
               & & & & & 5877.20 $\pm$   0.29 &  22.2 $\pm$  0.4 &   895.2 $\pm$   48.5 &  38.6 $\pm$  2.1  &  5877.52 $\pm$   0.29 &  21.8 $\pm$  0.4 &   860.8 $\pm$   53.7 &  37.6 $\pm$  2.3 \\ 
                 He{\,\sc ii} 6527.10 & 6530.34 $\pm$   15.56  & 10.8 $\pm$  38.4 &        --          &   3.0 $\pm$  10.0 &  6530.80 $\pm$   17.81 &  12.6 $\pm$  44.2  &  3.1 $\pm$  10.5 &   3.0 $\pm$  10.1 &  6531.23 $\pm$   17.46 &  12.9 $\pm$  43.3  &  3.3 $\pm$  10.6 &   3.2 $\pm$  10.3 \\ 
               & & & & & 6530.79 $\pm$   1.29 &  12.6 $\pm$  3.2 &    42.8 $\pm$   10.5 &   3.0 $\pm$  0.7  &  6531.22 $\pm$   1.27 &  12.9 $\pm$  3.1 &    45.3 $\pm$   10.6 &   3.2 $\pm$  0.7 \\ 
                 He{\,\sc ii} 6560.10 & 6564.26 $\pm$    3.44  & 20.2 $\pm$   9.1 &        --          &  34.7 $\pm$  16.4 &  6564.83 $\pm$    3.20 &  22.0 $\pm$   8.6  & 40.2 $\pm$  16.9 &  38.1 $\pm$  16.1 &  6565.19 $\pm$    3.25 &  22.9 $\pm$   8.9  & 42.5 $\pm$  17.9 &  40.5 $\pm$  17.0 \\ 
               & & & & & 6564.78 $\pm$   0.24 &  22.0 $\pm$  0.6 &   543.7 $\pm$   16.9 &  38.0 $\pm$  1.2  &  6565.13 $\pm$   0.24 &  22.9 $\pm$  0.7 &   575.0 $\pm$   17.8 &  40.4 $\pm$  1.3 \\ 
He{\,\sc i+ii} 6683.00, 6683.20 & 6683.28 $\pm$    9.34  & 31.0 $\pm$  26.2 &        --          &  28.1 $\pm$  28.2 &  6683.68 $\pm$    8.55 &  31.9 $\pm$  24.4  & 29.0 $\pm$  26.8 &  27.8 $\pm$  25.7 &  6684.21 $\pm$    8.45 &  31.3 $\pm$  23.9  & 28.4 $\pm$  26.1 &  26.9 $\pm$  24.7 \\ 
               & & & & & 6683.58 $\pm$   0.68 &  31.9 $\pm$  1.9 &   366.4 $\pm$   26.9 &  27.9 $\pm$  2.0  &  6684.11 $\pm$   0.67 &  31.3 $\pm$  1.9 &   358.6 $\pm$   26.1 &  27.0 $\pm$  2.0 \\ 
                  He{\,\sc i} 7065.19 & 7069.29 $\pm$   13.50  & 29.4 $\pm$  37.3 &        --          &  19.4 $\pm$  26.7 &  7069.45 $\pm$   11.81 &  31.7 $\pm$  33.0  & 21.5 $\pm$  24.6 &  20.7 $\pm$  23.6 &  7070.03 $\pm$   11.65 &  30.8 $\pm$  32.3  & 20.6 $\pm$  23.6 &  20.4 $\pm$  23.3 \\ 
               & & & & & 7069.35 $\pm$   1.16 &  31.6 $\pm$  3.2 &   219.0 $\pm$   24.6 &  20.6 $\pm$  2.3  &  7069.94 $\pm$   1.14 &  30.8 $\pm$  3.2 &   209.5 $\pm$   23.5 &  20.3 $\pm$  2.3 \\ 
                  N{\,\sc iv} 7109.35, 7122.98 & 7118.34 $\pm$    5.72  & 31.0 $\pm$  15.1 &        --          &  48.7 $\pm$  24.0 &  7119.19 $\pm$    4.85 &  33.9 $\pm$  12.9  & 57.2 $\pm$  22.2 &  55.0 $\pm$  21.3 &  7119.88 $\pm$    4.70 &  34.0 $\pm$  12.6  & 58.2 $\pm$  22.1 &  57.9 $\pm$  22.0 \\ 
      & & & & & 7119.07 $\pm$   0.49 &  33.9 $\pm$  1.3 &   567.5 $\pm$   22.2 &  54.8 $\pm$  2.1  &  7119.77 $\pm$   0.47 &  33.9 $\pm$  1.3 &   575.9 $\pm$   22.0 &  57.6 $\pm$  2.2 \\ 
                 He{\,\sc ii} 7177.52 & 7182.91 $\pm$   11.41  & 15.9 $\pm$  29.9 &        --          &   8.3 $\pm$  16.0 &  7181.31 $\pm$   15.46 &  19.7 $\pm$  42.0  &  7.0 $\pm$  15.9 &   6.7 $\pm$  15.3 &  7180.03 $\pm$   16.67 &  21.8 $\pm$  45.8  &  7.5 $\pm$  17.4 &   7.6 $\pm$  17.5 \\ 
                   & & & & & 7181.29 $\pm$   1.61 &  19.7 $\pm$  4.4 &    66.5 $\pm$   15.9 &   6.6 $\pm$  1.6  &  7180.02 $\pm$   1.74 &  21.8 $\pm$  4.8 &    72.0 $\pm$   17.4 &   7.5 $\pm$  1.8 \\ 
\hline
\enddata
\label{tbl:lines}
\end{deluxetable*}
\end{longrotatetable}

\subsection{WR classification }

 Visual spectral classification has been made using the emission line criteria
introduced by \cite{Smith-Shara-Moffat-96} for WN stars.  The He{\,\sc ii}
$\lambda$5411/He{\,\sc i} $\lambda$5875 flux ratio is about 0.82, which indicates
the spectral subtype is WN7. It should be noted that the fluxes are based on
Peak/Continuum values, and the absorption component is subtracted before
measuring the Peak value of the He{\,\sc i} $\lambda$5875 emission line. The FWHM
(He{\,\sc ii} $\lambda$4686) of 17.0 \AA\ and EW(He{\,\sc ii} $\lambda$5411) of
21.4 \AA, indicate WR 121-16 is a narrow-line WR star.  The
$\lambda\lambda$4200, 4541, and 5411 are pure He{\,\sc ii} lines in the Pickering
series, while $\lambda\lambda$4340 and 4861 are Pickering lines that might be
blended with H$\gamma$ and H$\beta$, respectively. Hydrogen will be detected,
if the fluxes of (H + He) $\lambda\lambda$4340 and 4861 lines clearly exceed
those of the pure He{\,\sc ii} lines. Taking the Peak/Continuum flux for
each line used in the calculation, $\lambda$4340/$\sqrt{\lambda4200 \times
\lambda4541}$ -1 and $\lambda$4861/$\sqrt{\lambda4541 \times \lambda5411}$ - 1
are -0.10 and -0.03, respectively, indicates this star has no detectable
hydrogen emission lines.  Furthermore, the simultaneous presence of C{\,\sc iv}
$\lambda$5808 and the even stronger N{\,\sc iv} $\lambda$7109 suggests that the
star is in the transition phase between WN and WC stages \citep{Shara-16}.
Hence this object can be classified as a WN7o/WC subtype star. 

As shown in Fig. 5 of \cite{Conti-Massey-89}, WN/C transition stars are
differentiated from WN stars in the diagram of the log of equivalent width of
C{\,\sc iv} $\lambda$5808 versus He{\,\sc ii} $\lambda$4686. This object has log
EW(He{\,\sc ii} $\lambda$4686) = 1.86 and log EW(C{\,\sc iv} $\lambda$5808) =
1.33, which are quite similar to those measured in the first WN/C star (1.83
and 1.38) found in M31 \citep{Shara-16}.

\begin{figure}
\includegraphics[width=\linewidth]{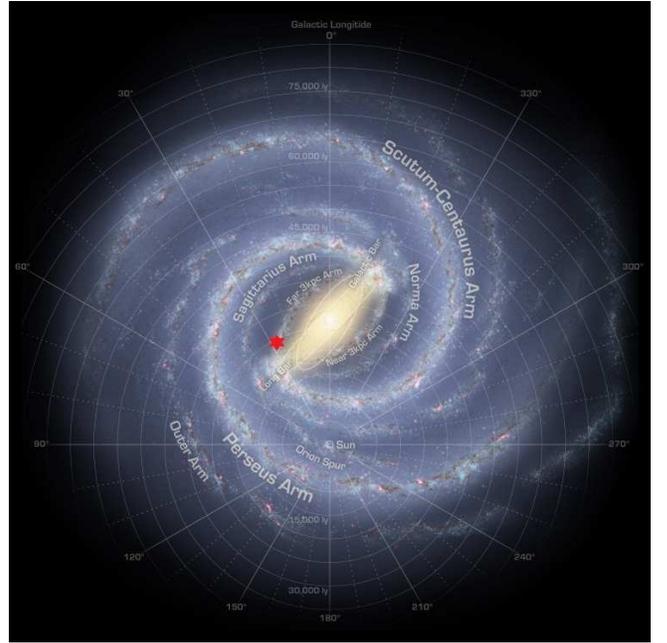} \caption{The
position of WR 121-16 labeled as a red star in the Galactic Plane. The
galactic coordinates are (\glon, \glat) = ($\rm 01^h51^m56\fs9$, $-02\arcdeg35\arcmin28\farcs0$). 
This object is located in the Far 3\,kpc Arm. It is about 7.11\,kpc from Earth, 
and 3.75\,kpc from the Galactic Center.  The background image is an artist's impression 
of the Milky Way, which shows the location of the spiral arms and the bulge.
Background image credit: NASA/JPL-Caltech/ESO/R. Hurt} 
\label{fig:position}
\end{figure}

\section{Stellar parameters}

\subsection{Distance}

The distance is necessary to derive absolute parameters such as luminosity,
stellar mass and mass loss rate. Thanks to \gaia, parallaxes and proper
motions are now available for 1.3 billion stars, which have been published
in the second \gaia\ data release \citep[GDR2,][]{GaiaCollaboration-18}.  As
discussed in \cite{Bailer-Jones-18} and \cite{Luri-18}, the conversion of
\gaia\ parallax \parallax\ to distance modifies the shape of the original
parallax probability distribution, when the parallax error \sigparallax\ is
larger than 0.1$\times$\parallax.  \cite{Bailer-Jones-18} have estimated
distances of 1.3 billion stars from their parallaxes, with the help of a Bayesian
approach. Using the distances given by this method, \cite{Sander-19}
re-examined a previously studied sample of WC stars to derive key properties of
the Galactic WC population. All quantities depending on the distance were
updated, while the underlying spectral analyses remain untouched. A similar
method has been applied to the Galactic WN population by \cite{Hamann-19}.
\cite{Rate-Crowther-19} obtained distances of 383 Galactic WR stars from GDR2
parallaxes, using the Bayesian method with a prior based on H{\,\sc ii} regions
and dust extinction.  Distances agree with those from \cite{Bailer-Jones-18}
for stars up to 2\,kpc from the Sun, though deviate thereafter due to differing
priors, leading to modest reductions in luminosities for recent WR
spectroscopic results. The parallax of WR 121-16 has been measured in GDR2 as
\parallax\ = 0.01 mas with a large error \sigparallax\ = 0.027 mas.  Hence, we
can not derive the distance of the star by the simple inversion of \dist =
1/\parallax. Fortunately, the distance to this star was estimated by
\cite{Bailer-Jones-18} as \dist = $7.11^{+1.56}_{-1.22}$\,kpc. We decide to use
this value in the following analyses. It's found that this star is right
located in the Far 3\,kpc Arm which was discovered by \cite{Dame-Thaddeus-08}.
Besides, this star is about 3.75\,kpc from the Galactic Center. The position of
this WN/C star is marked as a red star in Figure \ref{fig:position}.
 
\begin{figure*}
\centering
\includegraphics[scale=0.65]{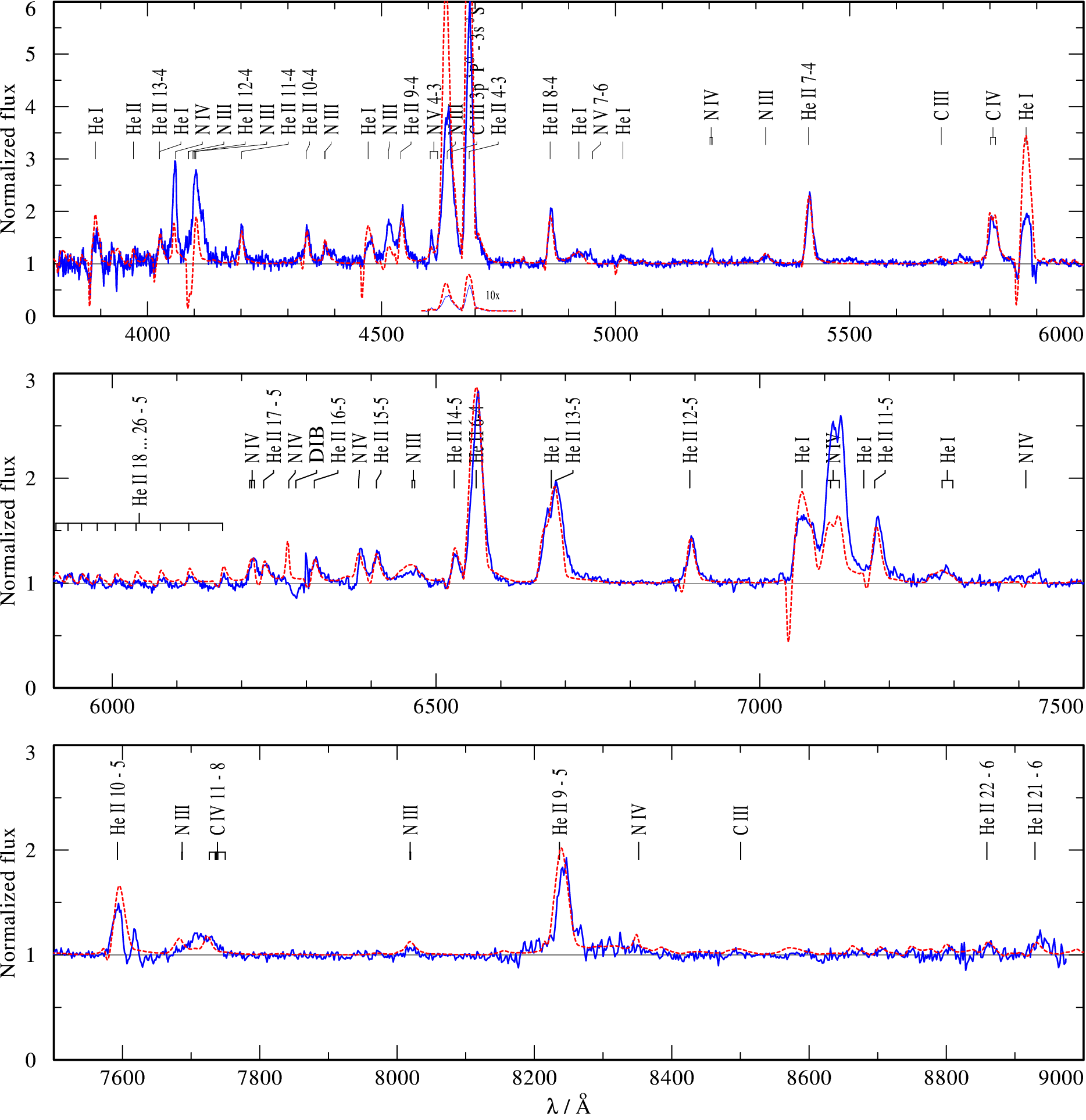}
\caption{Spectral fitting with the PoWR synthetic spectrum. The blue line is the normalized LAMOST spectrum, while the red line is the synthetic spectrum.}
\label{fig:model}
\end{figure*}

\subsection{PoWR model atmospheres}

The stellar parameters can be derived by comparing the observed spectra with
the synthetic ones calculated from the Potsdam Wolf-Rayet (PoWR) model
atmospheres, in which non-local thermal equilibrium (non-LTE), spherical
expansion and metal line blanketing are considered. Some grids of models with
fixed metallicity for WN stars \citep{Hamann-Grafener-04,Todt-15} and  WC stars
\citep{Sander-Hamann-Todt-12} can be publicly obtained from the PoWR models
website\footnote{http://www.astro.physik.uni-potsdam.de/PoWR.html}.

\begin{figure*}
\centering
\includegraphics[width=\textwidth]{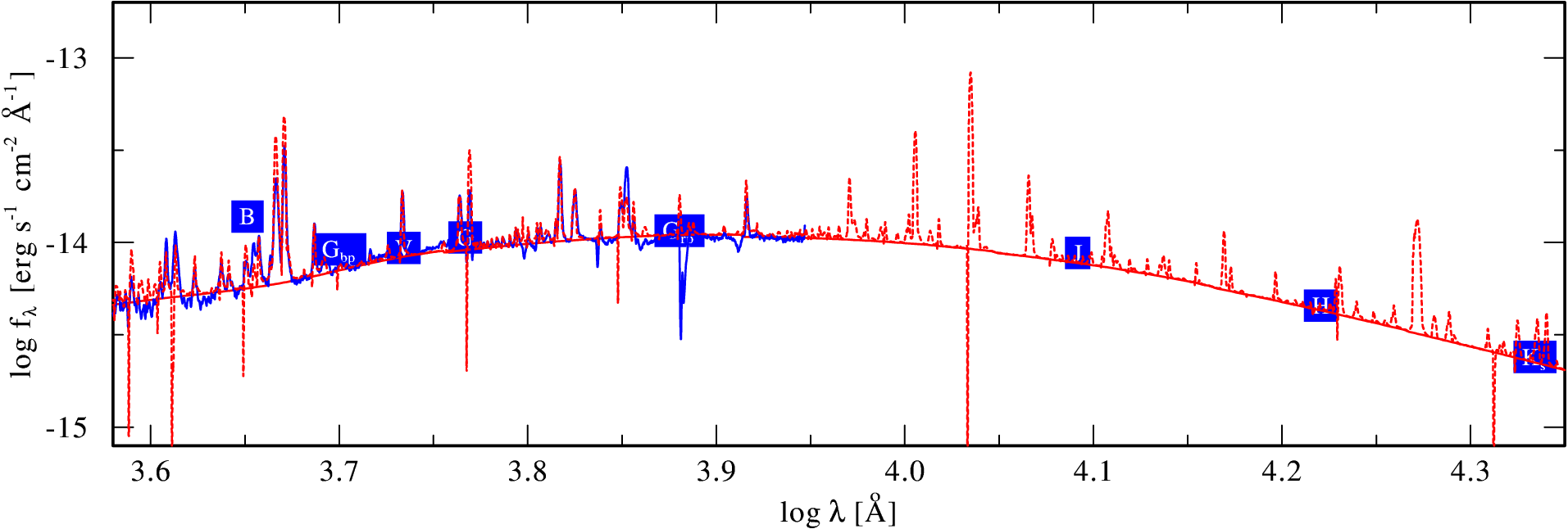}
\caption{Deriving the reddening $E(B - V)$ by comparing the synthetic spectrum (red dotted line) with the flux-calibrated 2.16\,m telescope spectrum (blue line) and with the observed \gaia\ and J, H, and K$_{\rm s}$ photometry. The continuum-only model flux (red line) is also shown for comparison.}
\label{fig:sed}
\end{figure*}

However, these models are not suitable for interpreting the transition WR
stars.  We changed the C, N, and O metallicity to generate a specific
synthetic spectrum for this WN/C star.
As the typical emission-line spectra of WR stars are
predominantly formed by recombination processes in their
dense stellar winds,  the continuum-normalized
spectrum shows a useful scale-invariance: for a given stellar
temperature $T_*$, which we define at the stellar radius
$R_*$ ($\tau_\mathrm{Rosseland}=20$),
and chemical composition, the equivalent
widths of the emission lines depend only on the ratio between
the volume emission measure of the wind and the area
of the stellar surface, to a first approximation. An equivalent
quantity, introduced by \citet{schmutz-89}, is the transformed radius
\begin{equation}
R_{\mathrm{t}} = 
R_* \left(\frac{v_\infty}{2500 \, \mathrm{km}\,\mathrm{s^{-1}}} 
\left/
\frac{\dot M \sqrt{D}}{10^{-4} \, M_\odot \, \text{
    yr}^{-1}} \right)^{2/3}
\right.
\end{equation}
with the terminal velocity $v_\infty$ and the microclumping parameter $D$,
which is defined as the density contrast between wind clumps and a smooth wind
of the same mass-loss rate.  Hence, we fitted first the normalized LAMOST
spectrum with this new PoWR  spectrum by eye and show the results in Figure
\ref{fig:model}. In a second step, we fitted the synthetic spectral energy
distribution to the observed flux-calibrated spectrum and photometry
(Figure~\ref{fig:sed}).

\begin{figure}
\centering
\includegraphics[width=\linewidth]{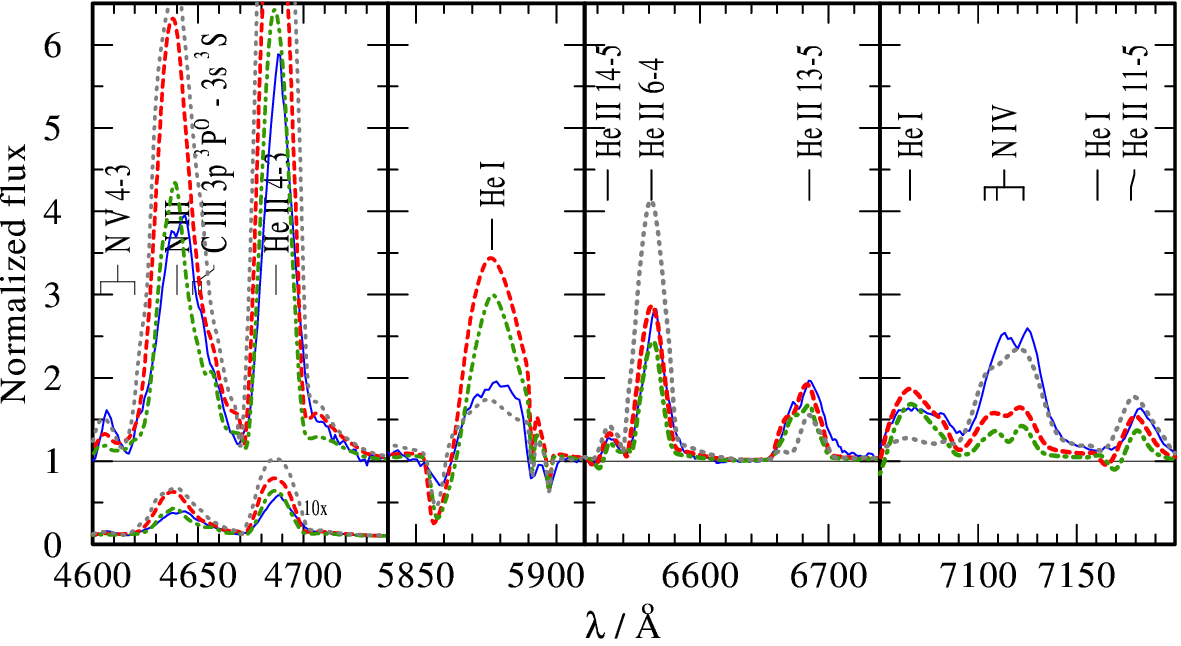}
\caption{Details from the optical spectrum: Observation (blue solid
  line) and our compromise model (red dashed). Also shown is a model
  with $T_*=42\,\rm kK$ and $\log(R_{\rm t}/R_\odot)=1.0$ (green
  dashed-dotted) and a model with  $T_*=56\,\rm kK$ and $\log(R_{\rm
    t}/R_\odot)=0.7$ (gray dotted). See text for details.
}
\label{fig:linefit}
\end{figure}

It appears difficult to achieve a good fit quality for
all observed optical emission lines.
The strongest lines N{\,\sc iii}
$\lambda$4634-$\lambda$4640 and He{\sc\,ii} $\lambda$4686 can be best
fitted with the stellar
temperature T$_* = 42\,$kK and the transformed radius log $R_t/R_\sun=1.0$,
while 
the fit is worse for all other emission lines (see Figure~\ref{fig:linefit}).
A better fit to the He{\sc\,i}  $\lambda5876$ and the N{\sc\,iv} lines can be obtained with
a model that has
$T_*$ = 56\,kK and log $R_t/R_\sun$ = 0.7
(see Figure~\ref{fig:linefit}). Our best compromise
fit that gives a sufficient
fit quality for almost all He{\sc\,ii} lines is therefore T$_*$ =
47$^{+9}_{-5}$\,kK and log $R_{\rm t}/R_\sun$ = 0.8$^{+0.2}_{-0.1}$. 

\begin{table}
\begin{center}
  \caption{PoWR parameters of the new WN/C star.}
    \begin{tabular}{lll}
\hline  
\hline  
T$_*$               & kK                     &  $47^{+9}_{-5}$           \\
log R$_t$           & R$_\odot$              &  $0.8^{+0.2}_{-0.1}$      \\
$v_\infty$          & km\,s$^{-1}$            &  1000$^{+200}_{-200}$     \\
log \.M             & M$_\odot$ yr$^{-1}$    & -4.97$^{+0.16}_{-0.20}$   \\     
R$_*$               & R$_\odot$              &  $4.14^{+1.4}_{-1.3}$   \\
log L               & L$_\odot$              &  4.88$^{+0.17}_{-0.15}$   \\
M$_*$               & M$_\odot$              &  7.1$^{+1.7}_{-1.1}$      \\  
D                   & clumping factor        &  4                        \\     
$X_\mathrm{H}$      & Mass fraction          & 0.0\%                     \\
$X_\mathrm{He}$     & Mass fraction          & 98\%                      \\
$X_\mathrm{Fe}$     & Mass fraction          & 0.14\%                    \\
$X_\mathrm{N}$      & Mass fraction          & 1.5$^{+1}_{-1}$\%         \\
$X_\mathrm{C}$      & Mass fraction          & 0.2$^{+0.1}_{-0.1}$\%     \\
$X_\mathrm{O}$      & Mass fraction          & $<$0.2\%                  \\
\hline  
    \end{tabular}
  \label{tbl:par}
\end{center}
\end{table}

Our compromise model fits all He{\sc\,ii} lines of the Pickering series well,
therefore we infer $X_{\rm H}$ = 0.0\% and $X_{\rm He}$ = 98\%. In the absence
of a UV spectrum required to derive the abundance of the iron group elements,
we adopted $X_{\rm Fe}$ = 0.14\%. The best fit to the emission lines of
C{\sc\,iii} and C{\sc\,iv} is for a carbon abundance of 0.2\% (by mass), that
is about 20 times the amount usually found in Galactic WNE and WNL stars
\citep[e.g.,][]{Hamann-06}, but consistent with the value derived by
\cite{Sander-Hamann-Todt-12} for other Galactic WN/C stars, like WR 58 (0.1\%)
or WR 126 (5\%).  As there seems to be no prominent oxygen lines in the
observations, we can only derive an upper limit of $X_{\rm O}$ $<$ 0.2\% from
the absence of the O{\sc\,iii} $\lambda$5592 line. We can not fit all the
lines of N{\sc\,iii}, N{\sc\,iv}, N{\sc\,v} simultaneously with one set of
stellar parameters, therefore it is harder to constrain the nitrogen abundance.
However, most lines can be fitted with the typical value for WN stars of 1.5\%,
which is also found for WR 58.

We use the clumping factor $D = 4$, because a model for a homogeneous wind
($D=1$) would result in too strong electron scattering line wings, and a higher
value of $D=10$ seems to underestimate the line wings.  We use a so-called
$\beta$-law to prescribe the velocity field $v(r) = v_\infty(1-R_*/r)^\beta$ in
the wind domain. The terminal velocity $v_{\infty}$ is estimated from the
widths of the emission lines, a value of 1000$^{+200}_{-200}$\,km\,s$^{-1}$
gives the best results.  The best fit is achieved with $\beta$ = 1.0, which is
commonly used for other Galactic WR stars.  However, the difficulties of
fitting the helium and nitrogen lines of different ionization stages
simultaneously might be resolved by a hydrodynamically consistent model, as
demonstrated by \citet{graefener-05} for the WC star WR 111. A Doppler
broadening of 100\,km\,s$^{-1}$ was applied to all the spectral lines and gives
a reasonable fit.
 
With the given \gaia\ parallax we fit our synthetic spectrum to the
flux-calibrated spectrum from the 2.16\,m telescope and to the observed \gaia\
and 2MASS (J, H, and K$_{\rm s}$) photometry in Figure \ref{fig:sed}. The best fit is
obtained for log $L/L_\sun$ = 4.88 and $E(B - V)\ = 1.20$\,mag, using the
extinction law by \cite{Fitzpatrick-99}.  We note that the luminosity is quite
low for a WR star, but WR 58 has also only a value of log $L/L_\sun$ = 4.95.
With this luminosity the mass-loss rate is log \.M = -4.97$^{+0.16}_{-0.20}\,M_\sun$ yr$^{-1}$.
The mass is estimated as about 7.1$^{+1.7}_{-1.1}$ $M_\sun$ based on the mass-luminosity
relation for WN stars by \cite{Grafener-11}. 
A visual absolute magnitude $M_V=-3.955$ is inferred for WR 121-16 from its distance and reddening.

\begin{figure}
\centering
\includegraphics[width=\linewidth]{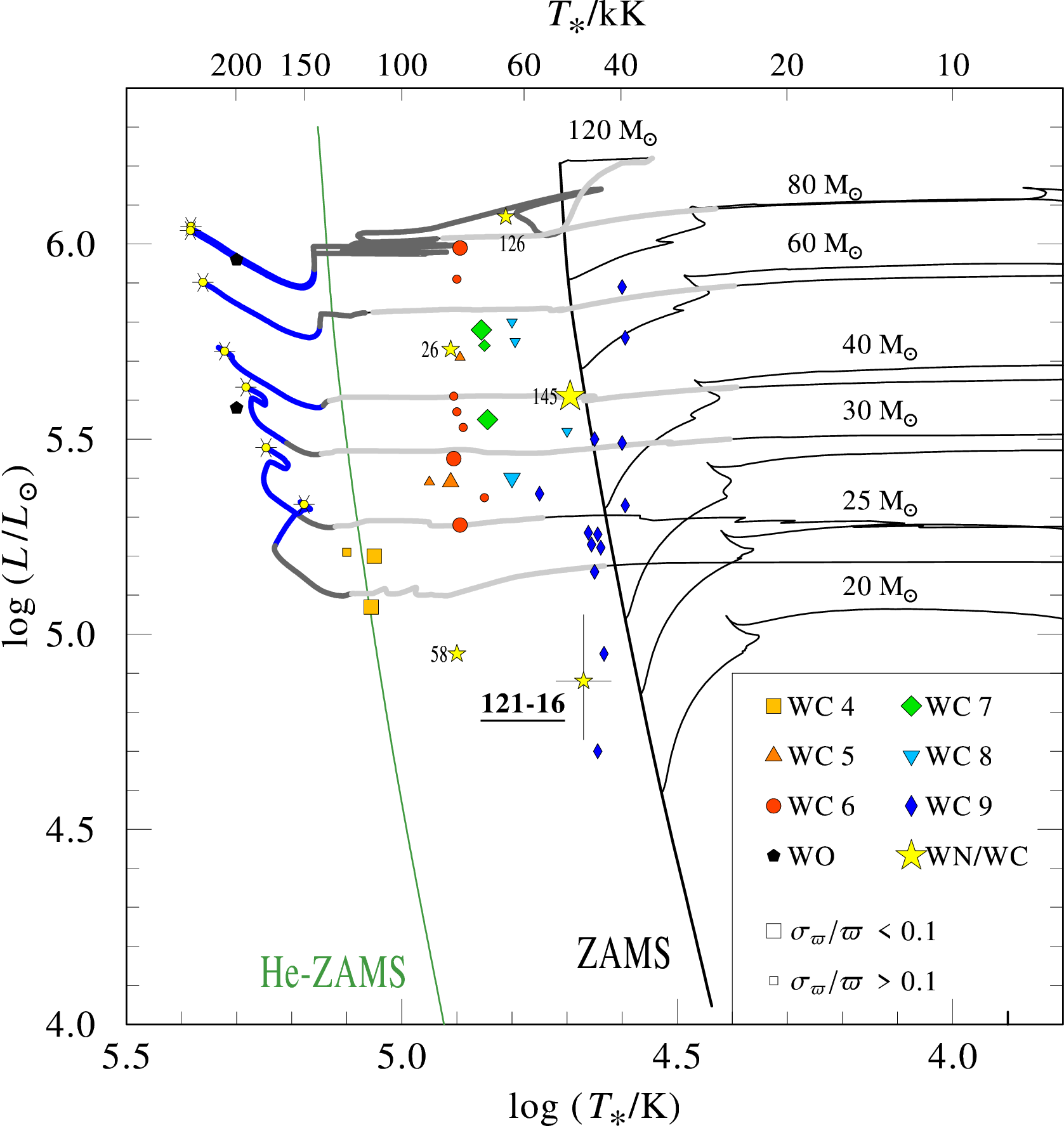}
\caption{HRD with the WC, WO and WN/C star positions compared to the evolutionary 
tracks from \cite{Chieffi-Limongi-13}. WR 121-16 is plotted as a yellow star with 
error bars. The thick lines indicate the WR phases of the tracks. This figure is adapted from \cite{Sander-19}.}
\label{fig:HRD}
\end{figure}

\subsection{Comparison with evolutionary tracks}

We show WR 121-16 and some WC, WO and four other WN/C stars in the
Hertzsprung-Russell diagram (HRD) in Figure \ref{fig:HRD}.  For comparison, the
evolutionary tracks from \cite{Chieffi-Limongi-13} obtained with the Frascati
Raphson Newton Evolutionary Code (FRANEC) are also shown in the same figure. WR
121-16 is located in the ``lower" luminosity regime. Even models with an
initial equatorial rotational velocity of  $v_{\rm rot,\,ini}$ = 300\,$\rm
km\,s^{-1}$ can not reach the position of WR 121-16. We will not discuss this
in detail as it is out of the scope of this paper.

\section{Is it a binary?}

The observed binary frequency (including the probable binaries) of the Galactic
WR stars is about 39\% \citep{vanderHucht-01}, and most of the binary
companions are O-type stars.  In WR+O binaries, the O-star's contribution often
dominates the overall composite spectrum \citep{Sander-Hamann-Todt-12}. No
signature of a possible O-type companion has been found in the spectrum of WR
121-16.  Moreover, there is another possibility that this star is a WN+WC
binary. However, there is no sign of radial velocity variation by comparing the
emission lines in the LAMOST spectrum with those in the 2.16\,m spectra.
Therefore, the spectrum of WR 121-16 is probably not composited of a WN+WC binary, but
rather a transition type WN/C star.

WR 121-16 has been collected in the American Association of Variable Star
Observers (AAVSO) International Variable Star Index
(VSX)\footnote{www.aavso.org/vsx/} with a name of ASASSN-V J185139.71-053451.1.
The magnitude varies between 13.95 and 14.14 in $V$-band but no period has been
found, and it was classified as Young Stellar Object (YSO) of unspecified
variable type \citep{Jayasinghe-18}.

The Renormalised Unit Weight Error (RUWE) is expected to be around 1.0, if the
astrometric observations of a source can be well fitted by the single-star
model. The RUWE may be significantly greater than 1.0, if the source is
non-single or there is problematic for the astrometric solution \citep{Belokurov-20}. Additionally,
a positive excess noise could indicate the source deviates from the standard
five-parameter astrometric model. For WR 121-16, RUWE = 1.05,
astrometric\_excess\_noise = 0.000 mas and astrometric\_excess\_noise\_sig =
0.000 provided by GDR2, indicate that this star is probably not a binary.

As a summary, the spectral and the time domain data, and the indices of
the astrometric solution all support that WR 121-16 is probably a single star.
Hence, the carbon and nitrogen emission lines are more likely from a WN/C
single star, rather than a composite spectrum from a WN+WC binary.

\section{Summary}

We have found a new Galactic WR star with coordinates of (R.A., Dec)$_{\rm J2000}$
= ($\rm 18^h51^m39\fs7$, $-05\arcdeg34\arcmin51\farcs1$) in the LAMOST testing
observations. According to the strong nitrogen and carbon emission lines
simultaneously showing in the optical spectra, WR 121-16 can be classified as a
transition WR of the WN7o/WC subtype.  With the help of \gaia\ parallax, the
distance from the Earth to WR 121-16 is inferred as $7.11^{+1.56}_{-1.22}$\,kpc,
and 3.75\,kpc from the Galactic Center.  Therefore, this object is located in
the Far 3\,kpc Arm.

By comparing the observed spectrum with the specific PoWR synthetic one,
physical parameters with a stellar temperature of 47$^{+9}_{-5}$\,kK, a mass
loss rate of $\dot{M}~=~10^{-4.97^{+0.16}_{-0.20}}~M_\sun~\rm yr^{-1}$, and a
stellar mass of 7.1$^{+1.7}_{-1.1}$ M$_\sun$ are derived. The carbon abundance
is 0.2\% by mass, which is about 20 times as high as that usually found in
Galactic WN stars. The luminosity is as low as log $L/L_\sun$ = 4.88$^{+0.17}_{-0.15}$, that is
similar to WR 58 which has a luminosity of log $L/L_\sun$ = 4.95.
Besides, WR 121-16 is more likely a transition WN/C type, rather than a WN+WC
binary.

\newpage
\acknowledgments 

The authors thank the anonymous referee for helpful comments that improved
this manuscript. W.Z. thanks Professor Paul Crowther for confirmation of the
name of WR 121-16 for this star. This work is supported by the Joint Research
Fund in Astronomy (No. U1531118) under cooperative agreement between the
National Natural Science Foundation of China (NSFC) and Chinese Academy of
Sciences (CAS).  This work is also supported by the NSFC (No. 11733006) and the
National Key R\&D Program of China grant (No.  2017YFA0402704). C.-H. Hsia
acknowledges the support from the Science and Technology Development Fund,
MacauSAR (file no.  0007/2019/A) and Faculty Research Grants of the Macau
University of Science and Technology (program no.  FRG-19-004-SSI).

The Guoshoujing Telescope (the Large Sky Area Multi-Object Fiber Spectroscopic
Telescope LAMOST) is a National Major Scientific Project built by the Chinese
Academy of Sciences.  Funding for the project has been provided by the National
Development and Reform Commission. LAMOST is operated and managed by the
National Astronomical Observatories, Chinese Academy of Sciences. We
acknowledge the support of the staff of the Xinglong 2.16\,m telescope. This
work was partially supported by the Open Project Program of the Key
Laboratory of Optical Astronomy, National Astronomical Observatories,
Chinese Academy of Sciences.

\newpage
\bibliography{ref}

\begin{thebibliography}{}
\expandafter\ifx\csname natexlab\endcsname\relax\def\natexlab#1{#1}\fi
\providecommand{\url}[1]{\href{#1}{#1}}
\providecommand{\dodoi}[1]{doi:~\href{http://doi.org/#1}{\nolinkurl{#1}}}
\providecommand{\doeprint}[1]{\href{http://ascl.net/#1}{\nolinkurl{http://ascl.net/#1}}}
\providecommand{\doarXiv}[1]{\href{https://arxiv.org/abs/#1}{\nolinkurl{https://arxiv.org/abs/#1}}}

\bibitem[{{Bailer-Jones} {et~al.}(2018){Bailer-Jones}, {Rybizki}, {Fouesneau},
  {Mantelet}, \& {Andrae}}]{Bailer-Jones-18}
{Bailer-Jones}, C.~A.~L., {Rybizki}, J., {Fouesneau}, M., {Mantelet}, G., \&
  {Andrae}, R. 2018, \aj, 156, 58, \dodoi{10.3847/1538-3881/aacb21}

\bibitem[{{Belokurov} {et~al.}(2020){Belokurov}, {Penoyre}, {Oh}, {Iorio},
  {Hodgkin}, {Evans}, {Everall}, {Koposov}, {Tout}, {Izzard}, {Clarke}, \&
  {Brown}}]{Belokurov-20}
{Belokurov}, V., {Penoyre}, Z., {Oh}, S., {et~al.} 2020, \mnras,
  \dodoi{10.1093/mnras/staa1522}

\bibitem[{{Blanton} \& {Roweis}(2007)}]{Blanton-07}
{Blanton}, M.~R., \& {Roweis}, S. 2007, \aj, 133, 734, \dodoi{10.1086/510127}

\bibitem[{{Breysacher} {et~al.}(1999){Breysacher}, {Azzopardi}, \&
  {Testor}}]{Breysacher-Azzopardi-Testor-99}
{Breysacher}, J., {Azzopardi}, M., \& {Testor}, G. 1999, \aaps, 137, 117,
  \dodoi{10.1051/aas:1999240}

\bibitem[{{Callingham} {et~al.}(2019){Callingham}, {Tuthill}, \&
  {Pope}}]{Callingham-Tuthill-Pope-19}
{Callingham}, J.~R., {Tuthill}, P.~G., \& {Pope}, B.~J.~S.~a. 2019, Nature
  Astronomy, 3, 82, \dodoi{10.1038/s41550-018-0617-7}

\bibitem[{{Chieffi} \& {Limongi}(2013)}]{Chieffi-Limongi-13}
{Chieffi}, A., \& {Limongi}, M. 2013, \apj, 764, 21,
  \dodoi{10.1088/0004-637X/764/1/21}

\bibitem[{{Conti} \& {Massey}(1989)}]{Conti-Massey-89}
{Conti}, P.~S., \& {Massey}, P. 1989, \apj, 337, 251, \dodoi{10.1086/167101}

\bibitem[{{Crowther}(2007)}]{Crowther-07}
{Crowther}, P.~A. 2007, \araa, 45, 177,
  \dodoi{10.1146/annurev.astro.45.051806.110615}

\bibitem[{{Crowther} {et~al.}(1998){Crowther}, {De Marco}, \&
  {Barlow}}]{Crowther-DeMarco-Barlow-98}
{Crowther}, P.~A., {De Marco}, O., \& {Barlow}, M.~J. 1998, \mnras, 296, 367,
  \dodoi{10.1046/j.1365-8711.1998.01360.x}

\bibitem[{{Crowther} {et~al.}(2003){Crowther}, {Drissen}, {Abbott}, {Royer}, \&
  {Smartt}}]{Crowther-03}
{Crowther}, P.~A., {Drissen}, L., {Abbott}, J.~B., {Royer}, P., \& {Smartt},
  S.~J. 2003, \aap, 404, 483, \dodoi{10.1051/0004-6361:20030503}

\bibitem[{{Crowther} \& {Smith}(1996)}]{Crowther-Smith-96}
{Crowther}, P.~A., \& {Smith}, L.~J. 1996, \aap, 305, 541

\bibitem[{{Crowther} {et~al.}(1995){Crowther}, {Smith}, \&
  {Willis}}]{Crowther-Smith-Willis-95}
{Crowther}, P.~A., {Smith}, L.~J., \& {Willis}, A.~J. 1995, \aap, 304, 269

\bibitem[{{Cui} {et~al.}(2012){Cui}, {Zhao}, {Chu}, {Li}, {Li}, {Zhang}, {Su},
  {Yao}, \& {et al.}}]{Cui-12}
{Cui}, X.-Q., {Zhao}, Y.-H., {Chu}, Y.-Q., {et~al.} 2012, Research in Astronomy
  and Astrophysics, 12, 1197, \dodoi{10.1088/1674-4527/12/9/003}

\bibitem[{{Dame} \& {Thaddeus}(2008)}]{Dame-Thaddeus-08}
{Dame}, T.~M., \& {Thaddeus}, P. 2008, \apjl, 683, L143, \dodoi{10.1086/591669}

\bibitem[{{Fitzpatrick}(1999)}]{Fitzpatrick-99}
{Fitzpatrick}, E.~L. 1999, \pasp, 111, 63, \dodoi{10.1086/316293}

\bibitem[{{Frei} \& {Gunn}(1994)}]{Frei-Gunn-94}
{Frei}, Z., \& {Gunn}, J.~E. 1994, \aj, 108, 1476, \dodoi{10.1086/117172}

\bibitem[{{Gaia Collaboration} {et~al.}(2018){Gaia Collaboration}, {Brown},
  {Vallenari}, {Prusti}, {de Bruijne}, {Babusiaux}, {Bailer-Jones}, {Biermann},
  \& {et al.}}]{GaiaCollaboration-18}
{Gaia Collaboration}, {Brown}, A.~G.~A., {Vallenari}, A., {et~al.} 2018, \aap,
  616, A1, \dodoi{10.1051/0004-6361/201833051}

\bibitem[{{G{\'o}mez-Gonz{\'a}lez} {et~al.}(2016){G{\'o}mez-Gonz{\'a}lez},
  {Mayya}, \& {Rosa-Gonz{\'a}lez}}]{Gomez-Gonzalez-Mayya-Rosa-Gonzalez-16}
{G{\'o}mez-Gonz{\'a}lez}, V.~M.~A., {Mayya}, Y.~D., \& {Rosa-Gonz{\'a}lez}, D.
  2016, \mnras, 460, 1555, \dodoi{10.1093/mnras/stw1118}

\bibitem[{{Gr{\"a}fener} \& {Hamann}(2005)}]{graefener-05}
{Gr{\"a}fener}, G., \& {Hamann}, W.~R. 2005, \aap, 432, 633,
  \dodoi{10.1051/0004-6361:20041732}

\bibitem[{{Gr{\"a}fener} {et~al.}(2011){Gr{\"a}fener}, {Vink}, {de Koter}, \&
  {Langer}}]{Grafener-11}
{Gr{\"a}fener}, G., {Vink}, J.~S., {de Koter}, A., \& {Langer}, N. 2011, \aap,
  535, A56, \dodoi{10.1051/0004-6361/201116701}

\bibitem[{{Hadfield} \& {Crowther}(2007)}]{Hadfield-Crowther-07}
{Hadfield}, L.~J., \& {Crowther}, P.~A. 2007, \mnras, 381, 418,
  \dodoi{10.1111/j.1365-2966.2007.12284.x}

\bibitem[{{Hamann} \& {Gr{\"a}fener}(2004)}]{Hamann-Grafener-04}
{Hamann}, W.~R., \& {Gr{\"a}fener}, G. 2004, \aap, 427, 697,
  \dodoi{10.1051/0004-6361:20040506}

\bibitem[{{Hamann} {et~al.}(2006){Hamann}, {Gr{\"a}fener}, \&
  {Liermann}}]{Hamann-06}
{Hamann}, W.~R., {Gr{\"a}fener}, G., \& {Liermann}, A. 2006, \aap, 457, 1015,
  \dodoi{10.1051/0004-6361:20065052}

\bibitem[{{Hamann} {et~al.}(2019){Hamann}, {Gr{\"a}fener}, {Liermann},
  {Hainich}, {Sander}, {Shenar}, {Ramachand ran}, {Todt}, \& {et
  al.}}]{Hamann-19}
{Hamann}, W.~R., {Gr{\"a}fener}, G., {Liermann}, A., {et~al.} 2019, \aap, 625,
  A57, \dodoi{10.1051/0004-6361/201834850}

\bibitem[{{Hillier}(1985)}]{Hillier-85}
{Hillier}, D.~J. 1985, \aj, 90, 1514, \dodoi{10.1086/113864}

\bibitem[{{Hillier} {et~al.}(1983){Hillier}, {Jones}, \&
  {Hyland}}]{Hillier-Jones-Hyland-83}
{Hillier}, D.~J., {Jones}, T.~J., \& {Hyland}, A.~R. 1983, \apj, 271, 221,
  \dodoi{10.1086/161189}

\bibitem[{{Jarrett} {et~al.}(2011){Jarrett}, {Cohen}, {Masci}, {Wright},
  {Stern}, {Benford}, {Blain}, {Carey}, {Cutri}, {Eisenhardt}, {Lonsdale},
  {Mainzer}, {Marsh}, {Padgett}, {Petty}, {Ressler}, {Skrutskie}, {Stanford},
  {Surace}, {Tsai}, {Wheelock}, \& {Yan}}]{Jarrett-11}
{Jarrett}, T.~H., {Cohen}, M., {Masci}, F., {et~al.} 2011, \apj, 735, 112,
  \dodoi{10.1088/0004-637X/735/2/112}

\bibitem[{{Jayasinghe} {et~al.}(2018){Jayasinghe}, {Kochanek}, {Stanek},
  {Shappee}, {Holoien}, {Thompson}, {Prieto}, {Dong}, \& {et
  al.}}]{Jayasinghe-18}
{Jayasinghe}, T., {Kochanek}, C.~S., {Stanek}, K.~Z., {et~al.} 2018, \mnras,
  477, 3145, \dodoi{10.1093/mnras/sty838}

\bibitem[{{Luo} {et~al.}(2015){Luo}, {Zhao}, {Zhao}, {Deng}, {Liu}, {Jing},
  {Wang}, {Zhang}, \& {et al.}}]{Luo-15}
{Luo}, A.~L., {Zhao}, Y.-H., {Zhao}, G., {et~al.} 2015, Research in Astronomy
  and Astrophysics, 15, 1095, \dodoi{10.1088/1674-4527/15/8/002}

\bibitem[{{Luri} {et~al.}(2018){Luri}, {Brown}, {Sarro}, {Arenou},
  {Bailer-Jones}, {Castro-Ginard}, {de Bruijne}, {Prusti}, \& {et
  al.}}]{Luri-18}
{Luri}, X., {Brown}, A.~G.~A., {Sarro}, L.~M., {et~al.} 2018, \aap, 616, A9,
  \dodoi{10.1051/0004-6361/201832964}

\bibitem[{{Massey} \& {Grove}(1989)}]{Massey-Grove-89}
{Massey}, P., \& {Grove}, K. 1989, \apj, 344, 870, \dodoi{10.1086/167854}

\bibitem[{{Massey} {et~al.}(2014){Massey}, {Neugent}, {Morrell}, \&
  {Hillier}}]{Massey-14}
{Massey}, P., {Neugent}, K.~F., {Morrell}, N., \& {Hillier}, D.~J. 2014, \apj,
  788, 83, \dodoi{10.1088/0004-637X/788/1/83}

\bibitem[{{Morgan} \& {Good}(1987)}]{Morgan-Good-87}
{Morgan}, D.~H., \& {Good}, A.~R. 1987, \mnras, 224, 435,
  \dodoi{10.1093/mnras/224.2.435}

\bibitem[{{Rate} \& {Crowther}(2019)}]{Rate-Crowther-19}
{Rate}, G., \& {Crowther}, P.~A. 2019, in The Gaia Universe, 45,
  \dodoi{10.5281/zenodo.3233991}

\bibitem[{{Roman-Lopes} {et~al.}(2020){Roman-Lopes},
  {Rom{\'a}n-Z{\'u}{\~n}iga}, {Borissova}, {Ram{\'\i}rez-Preciado},
  {Hern{\'a}ndez}, \& {Minniti}}]{Roman-Lopes-20}
{Roman-Lopes}, A., {Rom{\'a}n-Z{\'u}{\~n}iga}, C.~G., {Borissova}, J., {et~al.}
  2020, \apj, 891, 107, \dodoi{10.3847/1538-4357/ab72a6}

\bibitem[{{Rosslowe} \& {Crowther}(2015{\natexlab{a}})}]{Rosslowe-Crowther-15a}
{Rosslowe}, C.~K., \& {Crowther}, P.~A. 2015{\natexlab{a}}, \mnras, 449, 2436,
  \dodoi{10.1093/mnras/stv502}

\bibitem[{{Rosslowe} \& {Crowther}(2015{\natexlab{b}})}]{Rosslowe-Crowther-15b}
---. 2015{\natexlab{b}}, \mnras, 447, 2322, \dodoi{10.1093/mnras/stu2525}

\bibitem[{{Sander} {et~al.}(2012){Sander}, {Hamann}, \&
  {Todt}}]{Sander-Hamann-Todt-12}
{Sander}, A., {Hamann}, W.~R., \& {Todt}, H. 2012, \aap, 540, A144,
  \dodoi{10.1051/0004-6361/201117830}

\bibitem[{{Sander} {et~al.}(2019){Sander}, {Hamann}, {Todt}, {Hainich},
  {Shenar}, {Ramachandran}, \& {Oskinova}}]{Sander-19}
{Sander}, A.~A.~C., {Hamann}, W.~R., {Todt}, H., {et~al.} 2019, \aap, 621, A92,
  \dodoi{10.1051/0004-6361/201833712}

\bibitem[{{Schild} {et~al.}(1990){Schild}, {Smith}, \&
  {Willis}}]{Schild-Smith-Willis-90}
{Schild}, H., {Smith}, L.~J., \& {Willis}, A.~J. 1990, \aap, 237, 169

\bibitem[{{Schmutz} {et~al.}(1989){Schmutz}, {Hamann}, \&
  {Wessolowski}}]{schmutz-89}
{Schmutz}, W., {Hamann}, W.~R., \& {Wessolowski}, U. 1989, \aap, 210, 236

\bibitem[{{Shara} {et~al.}(2016){Shara}, {Miko{\l}ajewska}, {Caldwell},
  {I{\l}kiewicz}, {Drozd}, \& {Zurek}}]{Shara-16}
{Shara}, M.~M., {Miko{\l}ajewska}, J., {Caldwell}, N., {et~al.} 2016, \mnras,
  455, 3453, \dodoi{10.1093/mnras/stv2455}

\bibitem[{{Smith}(1968)}]{Smith-68}
{Smith}, L.~F. 1968, \mnras, 138, 109, \dodoi{10.1093/mnras/138.1.109}

\bibitem[{{Smith} {et~al.}(1990){Smith}, {Shara}, \&
  {Moffat}}]{Smith-Shara-Moffat-90}
{Smith}, L.~F., {Shara}, M.~M., \& {Moffat}, A. F.~J. 1990, \apj, 358, 229,
  \dodoi{10.1086/168978}

\bibitem[{{Smith} {et~al.}(1996){Smith}, {Shara}, \&
  {Moffat}}]{Smith-Shara-Moffat-96}
---. 1996, \mnras, 281, 163, \dodoi{10.1093/mnras/281.1.163}

\bibitem[{{Smith} {et~al.}(1994){Smith}, {Crowther}, \&
  {Prinja}}]{Smith-Crowther-Prinja-94}
{Smith}, L.~J., {Crowther}, P.~A., \& {Prinja}, R.~K. 1994, \aap, 281, 833

\bibitem[{{Su} \& {Cui}(2004)}]{Su-Cui-04}
{Su}, D.-Q., \& {Cui}, X.-Q. 2004, \cjaa, 4, 1, \dodoi{10.1088/1009-9271/4/1/1}

\bibitem[{{Todt} {et~al.}(2015){Todt}, {Sander}, {Hainich}, {Hamann}, {Quade},
  \& {Shenar}}]{Todt-15}
{Todt}, H., {Sander}, A., {Hainich}, R., {et~al.} 2015, \aap, 579, A75,
  \dodoi{10.1051/0004-6361/201526253}

\bibitem[{{van der Hucht}(2001)}]{vanderHucht-01}
{van der Hucht}, K.~A. 2001, \nar, 45, 135,
  \dodoi{10.1016/S1387-6473(00)00112-3}

\bibitem[{{Wang} {et~al.}(1996){Wang}, {Su}, {Chu}, {Cui}, \& {Wang}}]{Wang-96}
{Wang}, S.-G., {Su}, D.-Q., {Chu}, Y.-Q., {Cui}, X., \& {Wang}, Y.-N. 1996,
  \ao, 35, 5155, \dodoi{10.1364/AO.35.005155}

\bibitem[{{Zhao} {et~al.}(2012){Zhao}, {Zhao}, {Chu}, {Jing}, \&
  {Deng}}]{Zhao-12}
{Zhao}, G., {Zhao}, Y.-H., {Chu}, Y.-Q., {Jing}, Y.-P., \& {Deng}, L.-C. 2012,
  Research in Astronomy and Astrophysics, 12, 723,
  \dodoi{10.1088/1674-4527/12/7/002}

\bibitem[{{Zhao} {et~al.}(2018){Zhao}, {Fan}, {Ren}, {Ge}, {Zhang}, {Li},
  {Wang}, {Wang}, \& {et al.}}]{Zhao-18}
{Zhao}, Y., {Fan}, Z., {Ren}, J.-J., {et~al.} 2018, Research in Astronomy and
  Astrophysics, 18, 110, \dodoi{10.1088/1674-4527/18/9/110}

\end{thebibliography}

\label{lastpage}
\end{document}